\documentstyle{aipproc}

\def \beq{\begin{equation}}
\def \eeq{\end{equation}}
\def \beqn{\begin{eqnarray}}
\def \eeqn{\end{eqnarray}}
\def \cn{Collaboration}

\def \v#1#2{V_{#1#2}}

\begin{document}

\title{The Arrival of Charm\footnote{Enrico Fermi Institute report EFI-98-54,
November, 1998, hep-ph/9811359.  To be published in Proceedings of the Workshop
on Heavy Quarks at Fixed Target, Fermilab, Oct.~9--12, 1998.}} 

\author{Jonathan L. Rosner}
\address{Enrico Fermi Institute and Department of Physics \\
University of Chicago \\
5640 S. Ellis Avenue, Chicago IL 60637}

%\lefthead{LEFT head}
%\righthead{RIGHT head}
\maketitle

\begin{abstract}
Some of the theoretical motivations and experimental developments leading to
the discovery of charm are recalled.
\end{abstract}

% \leftline{\qquad PACS codes:  12.15.-y, 12.15.Mm, 14.40.Lb, 14.40.Gx}

\section{Introduction}

The discovery of charm was an exciting chapter in elementary particle physics.
The theoretical motivations were strong, the predictions were crisp, and the
experimental searches ranged from inadequate to serendipitous to inspired.
Perhaps we can learn something relevant to present-day searches from those
experiences.  I would like to describe the evolution of the case for charm,
some subsequent developments, and some questions which remain nearly a quarter
of a century later. 

The argument for charm was most compellingly made in the context of unification
of the weak and electromagnetic interactions, briefly described in Sec.~2.
Parallel arguments based on currents (Sec.~3) and quark-lepton analogies
(Sec.~4) also played a role, while gauge theory results (Sec.~5) strengthened
the case.  In the early 1970's, when electroweak theories began to be taken
seriously, theorists began to exhort their experimentalist colleagues in
earnest to seek charm (Sec.~6).  In the fall of 1974, the discovery of the
$J/\psi$ provided a candidate for the lowest-lying spin-triplet charm-anticharm
bound state, and several other circumstances hinted strongly at the existence
of charm (Sec.~7).  Nonetheless, not everyone was persuaded by this
interpretation, and it remained for open charm to be discovered before
lingering doubts were fully resolved (Sec.~8). 

Some progress in the post-discovery era is briefly noted in Sec.~9, while
some current questions are posed in Sec.~10.  A brief epilogue in Sec.~11 asks
whether the search for charm offers us any lessons for the future.  Part of the
author's interest in (recent) history stems from a review, undertaken with Val
Fitch, of elementary particle physics in the second half of the Twentieth
Century \cite{AIPIOP}, which is to be issued in a second edition in a year or
two. 

\section{Electroweak Unification}

The Fermi theory of beta decay \cite{EF} involved a pointlike interaction (for
example, in the decay $n \to p e^- \bar \nu_e$ of the neutron).  This
feature was eventually recognized as a serious barrier to its use in higher
orders of perturbation theory.  By contrast, quantum electrodynamics (QED),
involving photon exchange, was successfully used for a number of higher-order
calculations, particularly following its renormalization by Feynman, Schwinger,
Tomonaga, and Dyson \cite{SS}.

Attempts to describe the weak interactions in terms of particle exchange date
back to Yukawa \cite{Yuk}.  A theory of weak interactions involving exchange of
charged spin-1 bosons was written down by Oskar Klein in 1938 \cite{Klein}, to
some extent anticipating that of Yang and Mills \cite{YM} describing
self-interacting gauge particles. 

Once the $V-A$ theory of the weak interactions had been established in 1957
\cite{VA}, descriptions involving exchange of charged vector bosons were
proposed \cite{VB}.  These tried to unify charged vector bosons (eventually
called $W^\pm$) with the photon ($\gamma$) within a single SU(2) gauge
symmetry.  However, the (massless) photon couples to a vector current, while
the (massive) $W$'s couple to a $V-A$ current.  The SU(2) symmetry was
inadequate to discuss this difference.  Its extension by Glashow in 1961
\cite{SG} to an SU(2) $\times$ U(1) permitted the simultaneous description of
electromagnetic and charge-changing weak interactions at the price of
introducing a new {\it neutral} massive gauge boson (now called the $Z^0$)
which coupled to a specific mixture of $V$ and $A$ currents for each quark and
lepton. 

The Glashow theory left unanswered the mechanism by which the $W^\pm$ and $Z$
were to acquire their masses.  This was provided by Weinberg \cite{Wbg} and
Salam \cite{AS} through the Higgs mechanism \cite{Higgs}, whereby the SU(2)
$\times$ U(1) was broken spontaneously to the U(1) of electromagnetism.  Proofs
of the renormalizability of this theory, due in the early 1970's to G. 't
Hooft, M. Veltman, B. W. Lee, and J. Zinn-Justin \cite{renorm}, led to intense
interest in its predictions, including the existence of charge-preserving weak
interactions due to exchange of the hypothetical $Z^0$ boson.  By 1973, a
review by E. Abers and B. W. Lee \cite{AL} already was available as a guide to
searches for neutral weak currents and other phenomena predicted by the new
theory. 

\section{Currents}

Let $Q_l^{(+)}$ be the spatial integral of the time-component of the
charge-changing leptonic weak current, so that $Q_l^{(+)} |e^- \rangle_L =
|\nu_e \rangle_L$, where the subscript $L$ denotes a left-handed particle.  It
is a member of an SU(2) algebra, since it is just an isospin-raising operator.
Defining $Q_l^{(-)} = [Q_l^{(+)}]^\dag$, we can form $2 Q_{3l} \equiv
[Q_l^{(+)}, Q_l^{(-)}]$ and then find that the algebra closes: $[Q_{3l},
Q_l^{(\pm)}] = \pm Q_l^{(\pm)}$. In order to describe the decays of strange and
non-strange hadrons in a unified way with a suitably normalized weak hadronic
current, M. Gell-Mann and M. L\'evy \cite{GL} proposed in 1960 that the
corresponding hadronic charge behaved as $Q_h^{(+)} |n \cos \theta + \Lambda
\sin \theta \rangle = |p \rangle$, with $\sin \theta \simeq 0.2$.  Such a
current also is a member of an SU(2) algebra.  This allowed one to
simultaneously describe the apparent suppression of strange particle decay
rates with respect to strangeness-preserving weak interactions, and to account
for small violations of weak universality in beta-decay, which had become
noticeable as a result of radiative corrections \cite{Sirlin}. 

In 1963 N. Cabibbo adopted the idea of the Gell-Mann -- L\'evy current by
writing the weak current as
\beq
J^{\mu(+)} = \cos \theta J_{ \Delta S = 0}^{\mu (+)}
           + \sin \theta J_{ \Delta S = 1}^{\mu (+)}
\eeq
and using the newly developed flavor-SU(3) symmetry \cite{GMN} to evaluate
its matrix elements between meson and baryon states.  In the language
of the $u,d,s$ quarks this corresponded to writing the hadronic charge-changing
weak currents as
$$
Q_h^{(+)} = \left[ \begin{array}{c c c} 0 & \cos \theta & \sin \theta \\
                                      0 &      0      &      0      \\
                                      0 &      0      &      0      \\
\end{array} \right]~~,~~~Q_h^{(-)} = [Q_h^{(+)}]^\dag~~,~~~
$$
\beq
Q_{3h} \equiv \frac{1}{2}[Q_h^{(+)}, Q_h^{(-)}] =
\left[ \begin{array}{c c c}
1 &        0        &             0             \\
0 & - \cos^2 \theta & - \sin \theta \cos \theta \\
0 & - \sin \theta \cos \theta & - \cos^2 \theta \\
\end{array} \right]
\eeq
Again, the algebra closes: $[Q_{3h}, Q_h^{(\pm)}] = \pm Q_h^{(\pm)}$, so the
Cabibbo current is suitably normalized.  A good fit to weak semileptonic decays
of baryons and mesons was found in this manner, with $\sin \theta \simeq 0.22$.

As a student, I sometimes asked about the interpretation of $Q_{3h}$, which has
strangeness-changing pieces!  A frequent answer, reminiscent of the Wizard of
Oz, was:  ``Pay no attention to that [man behind the screen]!'' The neutral
current was supposed just to close the algebra, not to have  physical
significance. 

\section{Quark-Lepton Analogies:  Quartet Models}

Very shortly after the advent of the Cabibbo theory, a number of proposals
\cite{BjG,Hara,MO} sought to draw a parallel between the weak currents of
quarks and leptons in order to remove the strangeness-changing neutral currents
just mentioned.  Since the electron and muon each were seen to have their own
distinct neutrino \cite{Danby}, why shouldn't quarks be paired in the same way?
This involved introducing a quark with charge $Q=2/3$, carrying its own quantum
number, conserved under strong and electromagnetic but not weak interactions. 
As a counterpoise to the ``strangeness'' carried by the $s$ quark, the new
quantum number was dubbed ``charm'' by Bjorken and Glashow.  The analogy then
has the form
\beq
\left[ \begin{array}{c} \nu_e \\ e^- \end{array} \right]
\left[ \begin{array}{c} \nu_\mu \\ \mu^- \end{array} \right]
\Leftrightarrow
\left[ \begin{array}{c} u \\ d \end{array} \right]
\left[ \begin{array}{c} c \\ s \end{array} \right]~~~.
\eeq
The matrix elements of the hadronic $Q^{(+)}$ (we omit the subscript $h$) were
then 
\beq \label{eqn:mat}
\langle u|Q^{(+)}|d \rangle = \langle c|Q^{(+)}|s \rangle = \cos \theta~~,~~~
\langle u|Q^{(+)}|s \rangle = - \langle c|Q^{(+)}|d \rangle = \sin \theta~~~,
\eeq
while those of $Q_3$ were
\beq
\langle u|Q_3|u \rangle = \langle c|Q_3|c \rangle = -\langle d|Q_3|d \rangle =
-\langle s|Q_3|s \rangle = \frac{1}{2}~~~,
\eeq
with all off-diagonal ({\it flavor-changing}) elements equal to zero.  Here, as
before, $\sin \theta \simeq 0.22$.  Bjorken and Glashow were the first to call
the isospin doublet of non-strange charmed mesons ``$D$'' (for ``doublet''),
with $D^0 = c \bar u$ and $D^+ = c \bar d$.

\section{Gauge theory results}

The promotion of electroweak unification to a genuine gauge theory permitted
quantitative predictions of the properties of the fourth quark.

\subsection{The Glashow-Iliopoulos-Maiani (``GIM'') paper}

Taking his gauge theory of electroweak interactions seriously, Glashow in 1970
together with J. Iliopoulos and L. Maiani observed that the quartet model of
weak hadronic currents banished flavor-changing neutral currents to leading
order of momentum in higher orders of perturbation theory \cite{GIM}. Thus, for
example, higher-order contributions to $K^0$--$\bar K^0$ mixing, expected to
diverge in the $V-A$ theory or in a gauge theory without the charmed quark,
would now be cut off by $m_c$, where $m_c$ is the mass of the charmed quark. In
this manner an upper limit on the charmed quark mass of about 2 GeV was
deduced.  In view of the predominant coupling (\ref{eqn:mat}) of the charmed
quark to the strange quark, charmed particles should decay mainly to strange
particles, with a lifetime estimated to be about $\tau_{\rm charm} \simeq
10^{-13}$ s. 

The GIM paper contained a number of other specific predictions about the
properties of charmed particles.  Among these were: 

\begin{itemize}

\item A branching ratio of the charmed meson $D^0 = c \bar u$ to $K^- \pi^+$
of a few percent;

\item Strong production of charm-anticharm pairs;

\item Direct leptons in charm decays;

\item Charm production in neutrino reactions;

\item Neutral flavor-preserving currents;

\item The observability of a $Z^0$ in the direct channel of $e^+ e^-$
annihilations. 

\end{itemize}

These were all to be borne out over the next few years.  The discovery of the
$Z^0$ took longer, and was first made in a hadron rather than a lepton collider
\cite{Zz}. 

\subsection{Anomalies}

Once the electroweak theory was on firm theoretical grounds, it was noticed by
several authors in 1972 \cite{BIM,GG,GJ} that contributions to various
triangle diagrams involving fermion loops had to cancel.  For the electroweak
theory it was sufficient to consider the sum over all fermion species of
$I_{3L} Q^2$, where $I_{3L}$ is the weak isospin of the left-handed states and
$Q$ is their electric charge.  For the first family of quarks and leptons
the cancellation is arranged as follows:

\begin{center}
\begin{tabular}{l c c c c c}
Fermion: & $\nu_e$ & $e^-$ & $u$ & $d$ & Sum \\
Contribution: & $\frac{1}{2}(0)^2$ & $-\frac{1}{2}(-1)^2$ &
$\frac{1}{2}\left( 3 \right) \left(\frac{2}{3}\right)^2$ &
$-\frac{1}{2}\left( 3 \right) \left(-\frac{1}{3}\right)^2$ & 0 \\
Equal to: & 0 & $-\frac{1}{2}$ & $\frac{2}{3}$ & $-\frac{1}{6}$ & 0 \\
\end{tabular}
\end{center}

The corresponding cancellation for the second family reads
\beq
\nu_\mu + \mu + c + s = 0~~~,
\eeq
so that the charmed quark was {\it required} for such a cancellation, given the
existence of the muon and the strange quark.

\subsection{Rare kaon decays}

In a landmark 1973 paper, M. K. Gaillard and B. W. Lee \cite{GaL} took the
charmed quark seriously in calculating a host of processes involving kaons
to higher order in the new electroweak theory.  These included $K^0$--$\bar
K^0$ mixing and numerous rare decays such as $K_L \to (\mu^+ \mu^-,~
\gamma \gamma,~\pi^0 e^+ e^-,~\pi^0 \nu \bar \nu,~\ldots)$ and $K^+ \to
(\pi^+ e^+ e^-,~\pi^+ \nu \bar \nu,~\ldots)$.  The analyses of $K^0$--$\bar
K^0$ mixing and $K_L \to \mu^+ \mu^-$ indicated that $m_c^2 - m_u^2$ obeyed
a strong upper bound, while the failure of $K_L \to \gamma \gamma$ to be
appreciably suppressed indicated that $m_u^2 \ll m_c^2$.  Together these
results supported the GIM estimate of $m_c \le 2$ GeV and considerably
strengthened an earlier bound by Lee, J. Primack, and S. Treiman \cite{LPT}.

\section{Exhortations}

K. Niu and collaborators already had candidates for charm in emulsion as early
as 1971 \cite{Niu}.  These results, taken seriously by theorists in Japan
\cite{Ogawa,KM}, will be mentioned again presently.  Meanwhile, in the West,
theorists besides GIM began to urge their experimental colleagues to find
charm.  C. Carlson and P. Freund \cite{CF} discussed, among other things, the
properties of a narrow charm-anticharm bound state. George Snow \cite{Snow}
listed a number of features of charm production and decays.  Through an
interest in hadron spectroscopy, I became involved late in 1973 in these
efforts in collaboration with Gaillard and Lee. We started to look at charm
production and detection in hadron, neutrino, and electron-positron reactions.
It quickly became clear that a new quark, even one as light as 2 GeV, could
have been overlooked. 

Glashow spoke on charm at the 1974 Conference on Experimental Meson
Spectroscopy, held at Northeastern University \cite{EMS}.  In addition to the
properties mentioned in the earlier GIM paper, he told his experimental
colleagues to expect:

\begin{itemize}

\item Charm lifetimes ranging between $10^{-13}$ and $10^{-12}$ s;

\item Comparable branching ratios for semileptonic and hadronic decays;

\item An abundance of strange particles in the final state;

\item Dileptons in neutrino reactions (with the second lepton due to charm
decay).

\end{itemize}

He ended with the following charge to his colleagues:

\begin{quote} {\tt \centerline{WHAT TO EXPECT AT EMS-76}
\smallskip

There are just three possibilities:\\
1.  Charm is not found, and I eat my hat.\\
2.  Charm is found by hadron spectroscopers, and we celebrate.\\
3.  Charm is found by outlanders, and you eat your hats.}
\end{quote}

In the summer of 1974, Sam Treiman, then an editor of Reviews of Modern
Physics, pressed Ben Lee, Mary Gaillard, and me to write up our results with
the comment:  ``It's getting urgent.''  Our review of the properties of charmed
particles was eventually published in the April 1975 issue \cite{GLR}.  Better
late than never.  By then we were able to add an appendix dealing with the new
discoveries.  The body of our article (``GLR'') was written before them.  Our
conclusions, most of which I mentioned at a Gordon Conference late in the
summer of 1974, were as follows:

\begin{quote}{\tt

We have suggested some phenomena that might be indicative of charmed particles.
These include:\\

(a) ``direct'' lepton production,\\
(b) large numbers of strange particles,\\
(c) narrow peaks in mass spectra of hadrons,\\
(d) apparent strangeness violations,\\
(e) short tracks, indicative of particles with lifetime of order $10^{-13}$
sec.,\\
(f) di-lepton production in neutrino reactions,\\
(g) narrow peaks in $e^+e^-$ or $\mu^+ \mu^-$ mass spectra,\\
(h) transient threshold phenomena in deep inelastic leptoproduction,\\
(i) approach of the $(e^+ e^- \to {\tt hadrons})/(e^+ e^- \to \mu^+ \mu^-)$
ratio [``$R$''] to $3 \frac{1}{3}$, perhaps from above, and\\
(h) any other phenomena that may indicate a mass scale of 2 - 10 GeV.}
\end{quote}

A couple of these bear explanation.  ``Apparent strangeness violations''
can occur in the transitions $c \leftrightarrow d$; otherwise strangeness would
directly track charm (aside from a sign; the convention is that the strangeness
of an $s$ quark is $-1$, while the charm of a charmed quark is $+1$).  ``Narrow
peaks in $e^+e^-$ or $\mu^+ \mu^-$ mass spectra'' were not just dreamt up out
of the blue; we were aware of an effect in muon pairs at a mass around 3.5 GeV
\cite{oldJ} which could have been the lowest spin-triplet $c \bar c$ bound
state.  John Yoh remembers hearing this interpretation from Mary K. Gaillard
in the Fermilab cafeteria in August of 1974.  Our estimate of the width of this
state was about 2 MeV, based on extrapolating the Okubo-Iizuka-Zweig (OZI) rule
\cite{OZI} which suppressed ``hairpin'' quark diagrams. An early prediction by
T. Appelquist and H. D. Politzer \cite{AP} of the properties of $c \bar c$
bound states used QCD to anticipate a narrower spin-triplet than GLR.

I invited Glashow to the University of Minnesota in October of 1974 to speak on
charm and much else (including grand unified theories, which he was then
developing with Howard Georgi \cite{GUTs}). An unpersuaded curmudgeonly
astronomer turned to a younger colleague in the audience, whispering: ``When do
they bring in the men in white coats?'' The timing could not have been better.
Charm was to be discovered within a month.

\section{Hidden (and Not-So-Hidden) Charm}

As was suspected even before the days of QCD and asymptotic freedom, the ratio
$R \equiv \sigma(e^+ e^- \to {\rm hadrons})/\sigma(e^+ e^- \to \mu^+ \mu^-)$
probes the sum $\sum Q^2$ of the squared charges of quarks pair-produced at a
given c.m. energy.  Thus, above the resonances $\rho$, $\omega$, and $\phi$
which are features of low-energy $e^+ e^-$ annihilations into hadrons, one
expected to see $R = 3[(2/3)^2 + (-1/3)^2 + (-1/3)^2] = 2$, corresponding to
the three light quarks $u$, $d$, and $s$.  With wide errors, the ADONE Collider
at Frascati found this to be the case.  (See \cite{ADONE} for earlier
references.) 

In 1972 the Cambridge Electron Accelerator (CEA) was converted to an
electron-positron collider.  At energies above 3 GeV the cross section for $e^+
e^- \to {\rm hadrons}$, instead of falling with the expected $1/E_{\rm c.m.}^2$
behavior characteristic of pointlike quarks, was found to remain approximately
constant \cite{CEA}.  At $E_{\rm c.m.} = 4$ GeV, $R$ was $4.9 \pm 1.1$, while
it rose to $6.2 \pm 1.6$ at $E_{\rm c.m.} = 5$ GeV \cite{Richter}.  These
results were confirmed, with higher statistics, at the SPEAR machine
\cite{Richter}.  At the 1974 International Conference on High Energy Physics,
Burt Richter voiced concern about the validity of the naive quark
interpretation of $R$. 

The London Conference was distinguished by various precursors of charm in
addition to the rise in $R$ just mentioned.  Deep inelastic scattering of muon
neutrinos was occasionally seen (in about 1\% of events) to lead to a pair of
oppositely-charged muons.  One muon carried the lepton number of the incident
neutrino; the second could be the prompt decay product of charm. This
interpretation was mentioned by Ben Lee at the end of D. Cundy's rapporteur's
talk \cite{dimuons}.  Leptons produced at large transverse momenta \cite{lepts}
were due in part to prompt decays of charmed particles. John Iliopoulos
\cite{JI} not only laid out a number of the predictions for properties of
charmed particles, but bet anyone a case of wine that they would be discovered
by the next (1976) International Conference on High Energy Physics.  Though he
recalls several takers, they never paid off.

On November 11, 1974, the simultaneous discovery of the lowest-lying $^3S_1$
charm-anticharm bound state, with a mass of 3.1 GeV/$c^2$, was announced by
Samuel C. C. Ting and Burt Richter.  Ting's group, inspired in part by the
suggestion of a peak in an earlier experiment \cite{oldJ} and in part by an
innate confidence that lepton-pair spectra would reveal new physics, collided
protons produced at the Brookhaven Alternating-Gradient Synchrotron (AGS) with
a beryllium target to produce electron-positron pairs whose effective mass
spectrum was then studied with a high-resolution spectrometer \cite{J}.  The
new particle they discovered was called ``$J$'' (the character for ``Ting''
in Chinese).  Richter's group, working at SPEAR, wished to re-check anomalies
in the cross section for electron-positron annihilations that had shown up in
earlier running around a center-of-mass energy of 3 GeV. By carefully
controlling the beam energy, they were able to map out the peak of a narrow
resonance at 3.1 GeV \cite{psi}, which they called ``$\psi$'', a continuation
of the vector-meson series $\rho,\omega,\phi, \ldots$.  The dual name $J/\psi$
has been preserved. I was made aware of these discoveries by a call from Ben
Lee on November 11.  They certainly looked like charm to me, as well as to a
number of other people \cite{AP,Charms}. 

However, a large portion of the community offered alternative interpretations
\cite{PRLrefs}.  Some potential objections to charm (see the next Section) were
worth putting to experimental tests (e.g., by finding singly-charmed particles
\cite{JLRDPF}). However, I doubt the situation was ever as grave as implied by
the comment made to me in March of 1975 by Dick Blankenbecler at SLAC: 

\begin{quote}
{\tt Don't give up the ship.  It has just begun to sink.}
\end{quote}

\section{Open Charm}

In 1971, well before the discovery of the $J/\psi$, there were intimations of
particles carrying a single charmed quark through the short tracks they left in
emulsions, as studied by K. Niu and collaborators at Nagoya \cite{Niu}.  The
best candidate appears now to be an example of the rare decay $D^+ \to \pi^+
\pi^0$. Tony Sanda reminded us in this meeting \cite{Sanda} that by the 1975
International Conference on Cosmic Ray Physics this group had accumulated
\cite{ICRC} a significant sample of such ``short-lived particles.'' 

A candidate for the charmed baryon now called $\Lambda_c$ (as well as for
the decay $\Sigma_c \to \Lambda_c \pi$) was reported in neutrino interactions
in 1975 \cite{Lambdac}.  The properties of the $\Lambda_c$ and $\Sigma_c$ were
very close to those anticipated by an analysis of charmed-particle spectroscopy
\cite{DGG} which appeared shortly after the discussion of the $J/\psi$. 

Despite these indications, as well as the discovery of a candidate for
the first radial excitation (``$\psi'$'') of the $J/\psi$ \cite{psip} just 10
days after the observation of the $\psi$ in $e^+ e^-$ collisions, the charm
interpretation of the $J/\psi$ and $\psi'$ required several key tests to be
passed.

\subsection{Where was the $D \to \bar K \pi$ decay?}

The decays of charmed nonstrange mesons, with predicted masses of nearly 2
GeV/$c^2$, could involve a wide variety of final states, so that any individual
two-body (e.g., $D^0 \to K^- \pi^+$) or three-body (e.g., $D^+ \to K^- \pi^+
\pi^+$) mode should have a branching ratio of a few percent \cite{GIM}.

GLR attempted to estimate this effect using a current algebra model to estimate
multiple-pion production \cite{GLR}.  Unfortunately we used a value of the pion
decay constant $f_\pi$ high by $\sqrt{2}$ \cite{LQR}, and neglected other modes
besides $\bar K + n \pi$ \cite{EQ}.  Our results implied ${\cal B}(D^0 \to K
\pi)$ of nearly 50\% for a 2 GeV/$c^2$ charmed particle, clearly an
overestimate both in hindsight and intuitively (see, e.g., \cite{GIM}).  Our
result was quoted in the report \cite{NoD} of an initial SPEAR search which
failed to find charmed particles, and may have led to overconfidence in some
other proposed experiments \cite{NoC} which failed to find charm.  Subsequent
calculations (also taking into account non-zero pion mass), based both on the
current algebra matrix element and on a statistical model \cite{Fermi}, found
smaller $D \to \bar K \pi$ branching ratios than GLR \cite{LQR}. 

\subsection{Why did $R$ rise beyond its predicted value of $3 \frac{1}{3}$?}

The rise in $R$ observed at 4 GeV and higher was {\it too large} to account
for charm, which predicted $\Delta R = 3 Q_c^2 = 4/3$.  The resolution of this
problem was that pairs of $\tau$ leptons \cite{tau}, whose threshold
is $E_{\rm c.m.} = 2 m_\tau c^2 \simeq 3.56$ GeV, were also contributing to
$R$.  These $\tau$ leptons also diluted the rise in kaon multiplicity
expected above charm threshold.  This coincidence had all the aspects of a
mystery thriller \cite{HH}; the near-degeneracy of charm and $\tau$ production
thresholds is one of those effects (like the comparable masses of the
muon and pion) that seems just to have been put in to make the problem harder.

The value of $R$ is still a bit large in comparison with theoretical
expectations in the range covered by SPEAR \cite{Rval}.

\subsection{Where were the predicted electric dipole transitions from the
$\psi'$ to P-wave levels?}

The lowest P-wave charmonium levels (now called $\chi_c$) were predicted to lie
between the 1S and 2S levels \cite{Pmasses}.  Thus, one expected to be able
to see the electric dipole transitions $\psi' \to \gamma \chi_c$, leading to
monochromatic photons.  Initial inclusive searches using a NaI(Tl) detector at
SPEAR did not turn up these transitions \cite{Hof}, leading to some concern.

The problem turned out to be more experimentally demanding than originally
suspected.  By looking for the cascade transitions $\psi' \to \gamma \chi_c \to
\gamma \gamma J/\psi$, the DASP group, working at the DORIS storage ring at
DESY, presented the first results \cite{DASPchi} for the $\chi_{c1} = ~^3P_1$
level (with some possible admixture of $\chi_{c2} = ~^3P_2$).  By looking for
events of the form $\psi' \to \gamma \chi_c \to \gamma + (\pi \pi, K \bar K,
\ldots)$ and reconstructing the mass of the final hadronic state, the Mark I
group at SPEAR \cite{SPEARchi} detected states corresponding to both
$\chi_{c2}$ and $\chi_{c0} = ~^3P_0$. 

\subsection{Discovery of the $D$ mesons}

By 1975, estimates based on the mass of the $J/\psi$, on QCD \cite{DGG}, and on
potential models incorporating coupled-channel effects \cite{ECC} predicted $D$
masses in the range of 1.8 to 1.9 GeV/$c^2$, so that the rise in $R$ could, at
least in part, be accounted for by $D \bar D$ threshold.  Glashow urged Gerson
Goldhaber to re-examine the negative search results \cite{Riordan}. Together
with F. M. Pierre and other collaborators, Goldhaber incorporated
time-of-flight information to improve kaon identification, and found peaks in
$D^0 \to K^- \pi^+$ and $K^- \pi^+ \pi^- \pi^+$ \cite{Gh}, corresponding to a
mass which we now know to be 1.863 GeV/$c^2$. Low-multiplicity decays of the
$D^+$ were also seen shortly thereafter \cite{Dplus}. 

The first discoveries of $D$ mesons were announced in June of 1976.  This would
have been too late for the 1976 Meson Conference, which was traditionally held
in April, so Glashow could have lost his bet made at the 1974 Conference
\cite{EMS}.  (See, however, \cite{Lambdac}.) But meson spectroscopy was
entering a slower period, and the next conference was not held until 1977.
Since charm had clearly been discovered by outlanders, the participants were
obliged to eat their (candy) hats, graciously distributed by the conference
organizers. 

\subsection{The $\tau$ as interloper}

What about the $\tau$ lepton, whose appearance complicated the interpetation
of the SPEAR results?  It destroyed the anomaly cancellation, mentioned
earlier!  As a result, a new pair of quarks with charges 2/3 and $-1/3$, named
top and bottom by Harari \cite{HH}, had to be postulated.  Just such a quark
pair had been invented earlier (in 1973) by Kobayashi and Maskawa \cite{KM}
in order to explain the observed CP violation in kaon decays.  The discovery of
these quarks is another story, of which Fermilab has a right to be proud but
which we shall not mention further here.

\subsection{Total rate vs. purity in charm detection}

A question which arose in the search for charmed particles is being played out
again as present and future searches are planned. Is it better to work in a
relatively clean environment with limited rate, or in an environment where rate
is not a problem but backgrounds are high?  For charm in the mid-1970's, the
choice lay between the reaction $e^+ e^- \to \gamma^* \to c \bar c$,
contributing $\Delta R = 4/3$ above charm threshold, and fixed-target
proton-proton collisions at 400 GeV/$c^2$, with $\sigma_{c \bar c} = {\cal
O}(10^{-3}) \sigma_{\rm tot}$ but overall greater charm production rates than
in $e^+ e^-$ collisions. (The CERN Intersecting Storage Rings (ISR) were also
running at that time, providing proton-proton c.m. energies of up to 63 GeV but
with limited rates compared to fixed-target experiments.) 

After much time and effort, the balance eventually tipped in favor of fixed
target hadron (or photon) collisions.  (In photon collisions the photon can
couple directly to a charm-anticharm pair via the electric charge, leading to
diffractive production.)  Two advances that greatly enhanced the ability to
isolate charm were the use of the soft pion in $D^* \to D \pi$ decays
\cite{Nussinov} and the impressive growth in vertex detection technology
\cite{vertex}. 

\paragraph*{Soft pion tagging.}
The lowest-lying $^1S_0$ and $^3S_1$ bound states of a charmed quark and a
nonstrange antiquark are called $D$ and $D^*$, respectively.  Their masses are
such that $D^{*0}$ can decay to $D^0 \gamma$ and just barely to $D^0 \pi^0$,
while $D^{*+}$ can decay to $D^{*0} \gamma$ and just barely to $D^+ \pi^0$ or
$D^0 \pi^+$.  In the last case, the charged pion has a very low momentum with
respect to the $D^0$, and can be used to ``tag'' it.  One takes a hypothetical
set of $D^0$ decay products and combines them with the ``tagging'' pion.  If
the decay products really came from a $D^0$, the difference in effective masses
of the products with and without the extra pion should be $M(D^{*+}) - M(D^0)
\simeq 145$ MeV/$c^2$.  This method not only can help to see the $D^0$, but
can tell whether it was produced as a $D^0$ or a $\bar D^0$, since the only
low-mass combinations are $\pi^+ D^0$ or $\pi^- \bar D^0$.  This distinction
is important if one wishes to study $D^0$--$\bar D^0$ mixing or suppressed
decay modes of the $D^0$ (where the flavor of the decay products does not
necessarily indicate the flavor of the decaying state).

\paragraph*{Vertex detection.}
The earliest technique for detecting the short tracks made by charmed
particles, nuclear emulsions, was successfully used in Fermilab E-531 for the
detection of charmed particles produced in neutrino interactions, has been used
by Fermilab E-653 for the study of decays of charmed and $B$ mesons, and is
still in use for detecting decays of $\tau$ leptons produced in
neutrino-oscillation experiments \cite{CHORUS}. It has profited greatly from
automatic scanning methods introduced by Niu's group at Nagoya.  Still, it can
be subject to systematic errors, such as a bias against long neutral decay
paths. 

When it was realized that charmed particles could have lifetimes less than
$10^{-12}$ s, numerous attempts were made to improve the resolution of
existing devices such as bubble chambers and streamer chambers.  Some of these
are described in \cite{vertex}.

In the late 1970's, electronic spectrometers such as the OMEGA spectrometer at
CERN began to be equipped with new, high-resolution silicon vertex detectors.
These devices had the advantages of radiation hardness, excellent spatial
resolution, and electronic readout, making them {\it the} technique of choice
for resolving the tracks of short-lived particles in the busy environments of
hadro- and electroproduction.  Experiments which have profited from this
technique over the years include CERN WA-82, WA-89, WA-92 and Fermilab E-687,
E-691, E-769, E-791, and E-831 (FOCUS). 

\section{Examples of Further Progress}

\subsection{Emulsion results}

Emulsion studies of neutrino- and hadroproduction of charmed particles have
displayed the variation of lifetimes among charmed particles, measured the
decay constant $f_{D_s}$ of the charmed-strange meson $c \bar s \equiv D_s$,
and set limits on neutrino oscillations.  The scanning techniques pioneered by
the Nagoya group are beginning to be disseminated so that many institutions can
analyze future results. 

\subsection{Excited charmed mesons}

A meson containing a single heavy quark and a light antiquark is like a
hydrogen atom of the strong interactions.  The heavy quark corresponds to the
nucleus, while the antiquark (and its accompanying glue) correspond to the
electron and electromagnetic field.

The lowest orbitally excited states of charmed mesons follow an interesting
pattern rather different from that in charm-anticharm bound states.  In
$c \bar c$ levels, the charge-conjugation parity $C = (-1)^{L+S}$ prevents
the mixing of spin-singlet and spin-triplet levels with the same $L$.  Thus,
the properties of levels are best calculated by first coupling the $c$ and
$\bar c$ spins to $S=0$ or 1 and then coupling $S$ with the orbital angular
momentum $L$ to total angular momentum $J$.  One thus labels the states by
$^{2S+1}[L]_J$, where $[L] \equiv S,~P,~D,~F,~\ldots$ for $L = 0,~1,~2,~3,~
\ldots$.  In heavy-light states, however, nothing prevents mixing of $^1P_1$
and $^3P_1$ levels, and there is a favored pattern in the limit that the heavy
quark's mass approaches infinity \cite{DGG,JLRP,HQETP}.  One first couples the
light antiquark's spin $s = 1/2$ to the orbital angular momentum $L=1$ to
obtain the total angular momentum $j=1/2,~3/2$ carried by the light quark.  One
then couples $j$ to the heavy quark's spin $S_Q =1/2$ to obtain two pairs of
levels, as shown in Table 1. 

\begin{table}
\caption{Lowest orbitally-excited charmed mesons.}
\begin{center}
\begin{tabular}{c c c c c} \hline
$j$ & $J = j - \frac{1}{2}$ state & $J = j + \frac{1}{2}$ state & $l(D^{(*)}
\pi)$ & Width \\ 
\hline
1/2 & $? \to D \pi$ & $? \to D^* \pi~^a$ & 0 & Broad \\
3/2 & $D(2420) [\to D^* \pi$] & $D(2460) [\to D^{(*)} \pi$] & 2 & Narrow \\
\hline
\end{tabular}
\end{center}
\leftline{$^a$Candidate exists (see below).}
\end{table}

The $j=1/2$ states are expected to decay to $D^{(*)} \pi$ via S-waves and thus
to be broad and hard to find, while the $j=3/2$ states should decay via D-waves
and thus should be narrower and more easily distinguished from background.
The first orbitally excited charmed mesons were reported by the ARGUS
Collaboration \cite{ARP} in 1985.  Since then, considerable progress has been
made on these states by the ARGUS, CLEO, LEP, and fixed-target Fermilab
collaborations, with the properties of the $j=3/2$ states well mapped out.
There is now a candidate for a broad ($j=1/2$) state, with spin-parity $J^P =
1^+$, mass $M = 2.461^{+0.041} _{-0.034} \pm 0.010 \pm 0.032$ GeV, and width
$\Gamma = 290^{+101}_{-79} \pm 26 \pm 36$ MeV \cite{broad}. 

\subsection{Charmonium with antiprotons}

The ability to control the energy of an antiproton beam, first in the CERN
ISR \cite{ISR} and then in the Fermilab Antiproton Accumulator Ring \cite{ACC},
permitted the study of charmonium states through direct-channel production on
a gas-jet tartget.  A series of experiments studied the production and decay
of states like the $\eta_c$ (the $^1S_0$ charmonium ground state), the
$J/\psi$, and the $\chi_c$ levels, and led to the discovery of the $h_c$, the
$^1P_1$ level.  Precise measurements of masses and decay widths were made,
and an earlier claim \cite{etacp} for the $2^1S_0$ level, the $\eta_c'$, has
been disproved.  The search for this state, as well as for possible narrow
$c \bar c$ levels above $D \bar D$ threshold, continues at Fermilab as well
as elsewhere (see, e.g., \cite{DELetac}).

\subsection{Photo- and hadroproduction with vertex detection}

An impressive series of fixed-target experiments has refined the technique
of vertex detection using silicon strips or pixels \cite{vtxexps}, obtaining
unparalleled numbers of charmed particles.  Among the significant results are
detailed studies of lifetime differences among charmed particles, ranging
from greater than $10^{-12}$ s for the $D^+$ to less than $10^{-13}$ s
for the $\Omega_c = css$.

\subsection{Electron-positron collisions}

The ARGUS  and CLEO Collaborations continued to contribute significant results
on charmed particles produced in $e^+ e^-$ collisions, with results still
flowing from CLEO on such topics as the leptonic decay of the $D_s$ \cite{Ds}
and the spectroscopy of charmed baryons \cite{Charmedb}.

\section{Examples of Current Questions}

\subsection{Lifetime hierarchies}

The charmed-particle lifetimes mentioned in the previous Section, with
\beq 
\tau(\Omega_c) < \tau(\Lambda_c) < \tau(D^0) \simeq \tau(D_s) < \tau(D^+)
\eeq
varying by more than a factor of 10, continue to be a mild source of concern to
theorists.  The above hierarchy is better-understood \cite{Shifman,NS} than
that in strange particle decays, where lifetimes vary by more than a factor of
$600 \simeq \tau(K_L)/\tau(K_S)$.  However, the same methods which appear to
have described the charm lifetime hierarchy do not explain why $\tau(\Lambda_b)
/\tau(B^{+,0}) < 0.8$, whereas a ratio more like 0.9 to 0.95 is expected.  It
appears that non-perturbative effects, probably the main feature of the
lifetime differences for kaons and still important for charmed particles,
continue to have some residual effects even for the decays of the heavy $b$
quark.

\subsection{Decay constants}

The latest average for the $D_s$ decay constant \cite{Stone} is $f_{D_s} = 255
\pm 21 \pm 28$ MeV, based on observation of the decays $D_s \to \mu \nu,~\tau
\nu$.  We still need the values of the other heavy meson decay constants:
$f_D$, $f_B$, and $f_{B_s}$.  Lattice \cite{Draper} and QCD sum rule \cite{SN}
predictions for these quantities exist.  The value of $f_{D_s}$ is consistent
with predictions, though a bit on the high side.  The value of $f_D$ is in
principle accessible with present CLEO data samples \cite{LKGpc}.  One would
like to be able to distinguish between the quark-model prediction \cite{QMP}
$f_{D_s}/ f_D \simeq 1.25$ and the lattice/sum rule predictions of this ratio,
which range between 1.1 and 1.2. One may be able to isolate $D^+ \to \mu^+
\nu_\mu$ via the kinematics of the decay $D^{*+} \to \pi^0 D^+$ \cite{JRFD}. 

\subsection{Excited $D$ mesons}

Using heavy-quark symmetry, we can relate the properties of a meson containing
a heavy quark $Q$ and a light antiquark $\bar q$ to those where $Q$ is replaced
by another heavy quark $Q'$.  Thus, further study of excited $D = c \bar q$
mesons would give us information about the corresponding $\bar B = b \bar q$
states.  The properties of P-wave $b \bar q$ (``$B^{**}$'') mesons would be
very useful for ``tagging'' neutral $B$'s \cite{tags}, since a $\bar B^0$
resonates with a $\pi^-$ to form a $B^{**-}$ while a $B^0$ resonates with a
$\pi^+$ to form a $B^{**+}$. 

\subsection{Charm-anticharm mixing and CP violation}

Both mixing and CP-violating effects are expected to be far smaller for charmed
particles than for $B$'s \cite{charmCP}.  Since these effects are easier to
study in the charm system (at least in hadronic production, where charm
production is much easier than $b$ production), they are thus ideal for
displaying beyond-standard-model physics, since the standard-model effects
are so much smaller.

\section{Lessons?}

Should we be learning from history, or will we always be fighting the last war?
The search for charm has possible lessons, perhaps to be taken with a grain of
salt, for theory, experiment, and their synthesis in the form of future
searches. 

\subsection{Theory}

The optimism of theorists was justified in the search for charm.  The charmed
quark indeed was light, $m_c \simeq 1.5$ GeV/$c^2$.  Perturbative QCD was at
least a qualitative guide to the properties of charmonium and charmed
particles.  The discovery of the first quark with mass substantially exceeding
that of the QCD scale was a tremendous boost to the idea (already strongly
suggested by deep inelastic scattering) that fundamental quarks needed to be
taken seriously.

\subsection{Experiment}

Many searches for charmed particles were harder than people thought.  Sometimes
they were aided by sheer instrumental ``overkill,'' as in the case of the
superb mass resolution attained in the experiment which discovered the $J$
particle.  Sometimes the choice of a fortunate channel also helped, as in the
production of the $\psi$ by $e^+ e^-$ collisions with carefully controlled beam
energies, or in the choice of the $e^+ e^-$ decay mode in which to observe the
$J$.  Advances in instrumentation proved crucial, whether in the use of
particle identification to pull out the initial $D^0$ signal from background
or the study of charmed particles in high-background environments using vertex
detection.

\subsection{Future searches}

I do not see as clear a path in future searches as there was toward charm.  In
the case of supersymmetry, for example, the landscape looks very different.
There is a wide variety of predictions, and one is looking for the whole
supersymmetric system at once.  Alternate schemes for solving the problems
addressed by supersymmetry (e.g., dynamical electroweak symmetry breaking)
are not yet even formulated in a self-consistent manner. Perhaps that makes the
searches for physics beyond the standard model, which will be addressed in
future experiments here at Fermilab and elsewhere, even more exciting.

\section*{Acknowledgments}

I wish to thank Val Fitch for a pleasant collaboration on Ref.~\cite{AIPIOP},
Ikaros Bigi and Joel Butler for the chance to take this trip down memory lane,
and Peter Cooper, Mary K. Gaillard, John Iliopoulos, Scott Menary, Chris Quigg,
Tony Sanda, Lincoln Wolfenstein, and John Yoh for helpful comments.  This work
was supported in part the United States Department of Energy under Contract No.
DE FG02 90ER40560. 
\newpage

% Journal and other miscellaneous abbreviations for references
% Phys. Rev. D style
\def \ajp#1#2#3{Am.~J.~Phys.~{\bf#1}, #2 (#3)}
\def \apny#1#2#3{Ann.~Phys.~(N.Y.) {\bf#1}, #2 (#3)}
\def \app#1#2#3{Acta Phys.~Polonica {\bf#1}, #2 (#3)}
\def \arnps#1#2#3{Ann.~Rev.~Nucl.~Part.~Sci.~{\bf#1}, #2 (#3)}
\def \art{and references therein}
\def \cmp#1#2#3{Commun.~Math.~Phys.~{\bf#1}, #2 (#3)}
\def \cmts#1#2#3{Comments on Nucl.~Part.~Phys.~{\bf#1}, #2 (#3)}
\def \corn93{{\it Lepton and Photon Interactions:  XVI International
Symposium, Ithaca, NY August 1993}, AIP Conference Proceedings No.~302,
ed.~by P. Drell and D. Rubin (AIP, New York, 1994)}
\def \cp89{{\it CP Violation,} edited by C. Jarlskog (World Scientific,
Singapore, 1989)}
\def \dpff{{\it The Fermilab Meeting -- DPF 92} (7th Meeting of the
American Physical Society Division of Particles and Fields), 10--14
November 1992, ed. by C. H. Albright \ite~(World Scientific, Singapore,
1993)}
\def \dpf94{DPF 94 Meeting, Albuquerque, NM, Aug.~2--6, 1994}
\def \efi{Enrico Fermi Institute Report No. EFI}
\def \el#1#2#3{Europhys.~Lett.~{\bf#1}, #2 (#3)}
\def \flg{{\it Proceedings of the 1979 International Symposium on Lepton
and Photon Interactions at High Energies,} Fermilab, August 23--29, 1979,
ed.~by T. B. W. Kirk and H. D. I. Abarbanel (Fermi National Accelerator
Laboratory, Batavia, IL, 1979}
\def \hb87{{\it Proceeding of the 1987 International Symposium on Lepton
and Photon Interactions at High Energies,} Hamburg, 1987, ed.~by W. Bartel
and R. R\"uckl (Nucl. Phys. B, Proc. Suppl., vol. 3) (North-Holland,
Amsterdam, 1988)}
\def \ib{{\it ibid.}~}
\def \ibj#1#2#3{~{\bf#1}, #2 (#3)}
\def \ichep72{{\it Proceedings of the XVI International Conference on High
Energy Physics}, Chicago and Batavia, Illinois, Sept. 6--13, 1972,
edited by J. D. Jackson, A. Roberts, and R. Donaldson (Fermilab, Batavia,
IL, 1972)}
\def \ijmpa#1#2#3{Int.~J.~Mod.~Phys.~A {\bf#1}, #2 (#3)}
\def \ite{{\it et al.}}
\def \jmp#1#2#3{J.~Math.~Phys.~{\bf#1}, #2 (#3)}
\def \jpg#1#2#3{J.~Phys.~G {\bf#1}, #2 (#3)}
\def \ky85{{\it Proceedings of the International Symposium on Lepton and
Photon Interactions at High Energy,} Kyoto, Aug.~19-24, 1985, edited by M.
Konuma and K. Takahashi (Kyoto Univ., Kyoto, 1985)}
\def \Latt{{\it Lattice 98}, Proceedings, Boulder, CO, July 13--17, 1998}
\def \lkl87{{\it Selected Topics in Electroweak Interactions} (Proceedings
of the Second Lake Louise Institute on New Frontiers in Particle Physics,
15--21 February, 1987), edited by J. M. Cameron \ite~(World Scientific,
Singapore, 1987)}
\def \lnc#1#2#3{Lettere al Nuovo Cim.~{\bf#1}, #2 (#3)}
\def \lon{{\it Proceedings of the XVII International Conference on High Energy
Physics}, London, July 1974, edited by J. R. Smith (Rutherford Laboratory,
Chilton, England, 1974)}
\def \mpla#1#2#3{Mod.~Phys.~Lett.~A {\bf#1}, #2 (#3)}
\def \nc#1#2#3{Nuovo Cim.~{\bf#1}, #2 (#3)}
\def \nima#1#2#3{Nucl.~Instr.~Meth.~A {\bf#1}, #2 (#3)}
\def \np#1#2#3{Nucl.~Phys.~{\bf#1}, #2 (#3)}
\def \pisma#1#2#3#4{Pis'ma Zh.~Eksp.~Teor.~Fiz.~{\bf#1}, #2 (#3) [JETP
Lett. {\bf#1}, #4 (#3)]}
\def \pl#1#2#3{Phys.~Lett.~{\bf#1}, #2 (#3)}
\def \plb#1#2#3{Phys.~Lett.~B {\bf#1}, #2 (#3)}
\def \ppmsj#1#2#3{Proc.~Phys.-Math.~Soc.~Japan {\bf#1}, #2 (#3)}
\def \pr#1#2#3{Phys.~Rev.~{\bf#1}, #2 (#3)}
\def \pra#1#2#3{Phys.~Rev.~A {\bf#1}, #2 (#3)}
\def \prd#1#2#3{Phys.~Rev.~D {\bf#1}, #2 (#3)}
\def \prl#1#2#3{Phys.~Rev.~Lett.~{\bf#1}, #2 (#3)}
\def \prp#1#2#3{Phys.~Rep.~{\bf#1}, #2 (#3)}
\def \ptp#1#2#3{Prog.~Theor.~Phys.~{\bf#1}, #2 (#3)}
\def \rmp#1#2#3{Rev.~Mod.~Phys.~{\bf#1}, #2 (#3)}
\def \rp#1{~~~~~\ldots\ldots{\rm rp~}{#1}~~~~~}
\def \si90{25th International Conference on High Energy Physics, Singapore,
Aug. 2-8, 1990}
\def \slc87{{\it Proceedings of the Salt Lake City Meeting} (Division of
Particles and Fields, American Physical Society, Salt Lake City, Utah,
1987), ed.~by C. DeTar and J. S. Ball (World Scientific, Singapore, 1987)}
\def \slac89{{\it Proceedings of the XIVth International Symposium on
Lepton and Photon Interactions,} Stanford, California, 1989, edited by M.
Riordan (World Scientific, Singapore, 1990)}
\def \smass82{{\it Proceedings of the 1982 DPF Summer Study on Elementary
Particle Physics and Future Facilities}, Snowmass, Colorado, edited by R.
Donaldson, R. Gustafson, and F. Paige (World Scientific, Singapore, 1982)}
\def \smass90{{\it Research Directions for the Decade} (Proceedings of the
1990 Summer Study on High Energy Physics, June 25 -- July 13, Snowmass,
Colorado), edited by E. L. Berger (World Scientific, Singapore, 1992)}
\def \stone{{\it B Decays}, edited by S. Stone (World Scientific,
Singapore, 1994)}
\def \tasi90{{\it Testing the Standard Model} (Proceedings of the 1990
Theoretical Advanced Study Institute in Elementary Particle Physics,
Boulder, Colorado, 3--27 June, 1990), edited by M. Cveti\v{c} and P.
Langacker (World Scientific, Singapore, 1991)}
\def \Vanc{XXIX International Conference on High Energy Physics, Vancouver,
BC, Canada, July 23--29, 1998, Proceedings}
\def \yaf#1#2#3#4{Yad.~Fiz.~{\bf#1}, #2 (#3) [Sov.~J.~Nucl.~Phys.~{\bf #1},
#4 (#3)]}
\def \zhetf#1#2#3#4#5#6{Zh.~Eksp.~Teor.~Fiz.~{\bf #1}, #2 (#3) [Sov.~Phys.
- JETP {\bf #4}, #5 (#6)]}
\def \zp#1#2#3{Zeit.~Phys.~{\bf#1}, #2 (#3)}
\def \zpc#1#2#3{Zeit.~Phys.~C {\bf#1}, #2 (#3)}

\end{document}